\documentclass[prl,aps,twocolumn,floatfix,superscriptaddress]{revtex4}

\usepackage{amssymb,amsmath,amstext}               
\usepackage{graphicx}
\usepackage{bm}                                    
\usepackage{appendix}                              
\usepackage[utf8]{inputenc}
\usepackage{ulem}
\usepackage{latexsym}
\usepackage{url}
\usepackage[colorlinks=true,citecolor=blue,linkcolor=magenta,urlcolor=blue]{hyperref}

\usepackage[x11names,svgnames,dvipsnames]{xcolor}

\newcommand{\tsc}{TSC}
\newcommand{\qhe}{QHE}

\def\be{\begin{equation}}
\def\ee{\end{equation}}
\def\bea{\begin{eqnarray}}
\def\eea{\end{eqnarray}}
\def\ra{\rangle}
\def\la{\langle}
\def\bi{\begin{itemize}}
\def\ei{\end{itemize}}
\def\ben{\begin{enumerate}}
\def\een{\end{enumerate}}

\definecolor{dgreen} {RGB}{78,138,21}

\begin{document} 

\title{Six-dimensional time-space crystalline structures}

\author{Giedrius \v{Z}labys} 
\affiliation{Institute of Theoretical Physics and Astronomy, Vilnius University, Saul\.{e}tekio 3, LT-10257 Vilnius, Lithuania}
\author{Chu-hui Fan} 
\affiliation{Instytut Fizyki Teoretycznej, 
Uniwersytet Jagiello\'nski, ulica Profesora Stanis\l{}awa \L{}ojasiewicza 11, PL-30-348 Krak\'ow, Poland}
\author{Egidijus Anisimovas} 
\affiliation{Institute of Theoretical Physics and Astronomy, Vilnius University, Saul\.{e}tekio 3, LT-10257 Vilnius, Lithuania}
\author{Krzysztof Sacha} 
\affiliation{Instytut Fizyki Teoretycznej, 
Uniwersytet Jagiello\'nski, ulica Profesora Stanis\l{}awa \L{}ojasiewicza 11, PL-30-348 Krak\'ow, Poland}

\begin{abstract}
Time crystalline structures are characterized by regularity that single-particle or many-body systems manifest in the time domain, 
closely resembling the spatial regularity of ordinary space crystals. Here we show that time and space crystalline 
structures can be combined together and even six-dimensional time-space lattices can be realized. As an example, 
we demonstrate that such time-space crystalline structures can reveal the six-dimensional quantum Hall effect quantified by 
the third Chern number. 
\end{abstract}
\date{\today}

\maketitle

Ordinary space crystals are characterized by a spatially periodic distribution of atoms observed at a fixed instance of time, 
i.e.\ the moment of the experimental detection. This periodic distribution presents a manifestation of spontaneous symmetry 
breaking, whereby continuous translational symmetry of the underlying Hamiltonian is disrespected and narrowed down to the discrete
translational symmetry of a crystal.
In time crystals \cite{Shapere2012,Wilczek2012,Sacha2017rev,khemani2019brief,Guo2020,SachaTC2020}, 
the roles of time and space are inverted. One fixes the position in space, which corresponds to the location of a detector, 
and asks if the probability of clicking of the detector behaves periodically in time. It was shown that discrete time crystal 
behavior can emerge spontaneously in a many-body system
\cite{Sacha2015,Khemani16,ElseFTC,Huang2017,Zhang2017,Choi2017,Russomanno2017,Giergiel2018a,
Surace2018,Yu2018,Giergiel2018c,Pal2018,Smits2018,
Rovny2018,Autti2018,Kosior2018,Matus2019,
Pizzi2019,Pizzi2019a,Kozin2019,
Giergiel2020,Kuros2020,Autti2020,Wang2020}; here, the interacting many-body system conspires
to support structures that have lower symmetry (larger period) than the driving signal. However, in time-crystal research the
phenomenon of spontaneous symmetry breaking is not the only interesting scenario. Crystalline structures in time can also be engineered by suitable
external time-periodic driving \cite{Guo2013,sacha15a,Sacha2017rev,Guo2020,SachaTC2020}. 
In the latter case we deal with a situation similar to photonic crystals \cite{Joannopoulos_Book} where periodic modulation of the refractive index in space has to be imposed externally. Although no spontaneous symmetry breaking occurs, the situation is 
nonetheless interesting. Here, one focuses on the time-periodic behavior that is described by temporal counterparts of 
crystalline solid-state models.
Analogs of various condensed-matter phases have been already investigated in time crystals: Anderson and many-body localization, Mott insulator and topological phases can be observed in the time domain \cite{sacha15a,sacha16,delande17,Mierzejewski2017,
Giergiel2018,Lustig2018,Giergiel2018b}. 
Periodically driven systems can also reveal crystalline structures in the phase space \cite{Guo2013,Guo2016,Guo2016a,Liang2017}. See \cite{Sacha2017rev,Guo2020,SachaTC2020} for recent reviews.

In the present letter, we demonstrate the notion of \emph{time-space crystals} (\tsc) that support both spatial and temporal periodic structures simultaneously \cite{Li2012,Gomez-Leon2013,Messer2018,Fujiwara2019,Cao2020,Chakraborty2020,Martinez2020,Gao2020}. We begin with a single particle moving in a one-dimensional (1D) spatially periodic potential. If such a potential is periodically and resonantly driven in time, a time crystalline structure can be created and combined with the periodic structure in space to form a 2D \tsc. 
Similar shaking can be realized in all three orthogonal directions and allows one to create a 6D \tsc. Realization of a 6D crystalline structure paves the way towards investigation of 6D condensed-matter phases. 
Here we demonstrate how these structures can be endowed with synthetic gauge fields \cite{Jaksch2003,Gerbier2010,kolovsky11,AidelsburgerEtAl2013,miyake13harper,Goldman2014,Eckardt16review}, and this development completes the toolbox necessary for the realization of the 6D quantum Hall effect (\qhe). Here, in addition to the first Chern number \cite{ThoulessEtAl1982}, the nonlinear quantized response is characterized by the third Chern number \cite{LeeEtAl2018,PetridesEtAl2018}. 
Our work complements and extends several previous proposals \cite{Price15PRL,Price17PRA}. Ref.~\cite{Price15PRL} focused on the 4D \qhe\ \cite{LohseEtAl2018} realized by introducing a single extra dimension to a 3D lattice and Ref.~\cite{Price17PRA} sought to access 6D physics by reinterpreting the energy levels of a harmonic trap as a synthetic dimension.
 From the perspective of accessing higher-dimensional physics with the help of extra or synthetic dimensions \cite{Boada12PRL,CeliEtAl2014,ManciniEtAl2015,StuhlEtAl2015,LiviEtAl2016} our work seeks to promote the time as a resource suitable for doubling of the number of dimensions \cite{Sacha2017rev,Peng2018,Guo2020,SachaTC2020}.

{\it 2D time-space crystals.}---Let us start with a particle in a 1D spatially periodic potential which is periodically shaken in time. For ultracold atoms, such a potential can be realized by modulating two counter-propagating laser beams \cite{Eckardt16review}. Switching to the frame moving with the lattice the scaled Hamiltonian (in the recoil units \cite{sup1}) reads
\be
  H(x,p_x,t) = p_x^2 + V_0\sin^2x + p_x\lambda\omega\sin\omega t.
\label{H_basic}
\ee
Here $V_0$ is the depth of the optical lattice, and the last term in (\ref{H_basic}) results from the transition to the moving frame, see \cite{sup1}. The position of the optical lattice is harmonically modulated with a small amplitude $\lambda$ and frequency $\omega$ which is chosen resonant with the motion of a particle. Classically, this means that the particle of energy $E < V_0$ is moving periodically in a single lattice site with the frequency $\Omega(E) = \omega / s$, with an integer $s$. 
In order to understand how time crystalline structure is created let us apply the classical secular approximation. First, we introduce the canonical variables where the position of the particle on a resonant orbit is described by the angle $\theta \in [-\pi,\pi)$, and the corresponding conjugate momentum is the action $I$. Then, we switch to the frame moving along the orbit, $\Theta = \theta - \omega t/s$, and finally average the resulting Hamiltonian over time. This yields the time-independent effective Hamiltonian, $H_{i} = P^2/2m_{\rm eff} - V_{\rm eff} \cos (s\Theta)$ 
that describes the motion of a particle in the $i$-th lattice site in the vicinity of the resonant orbit characterized by the resonant value of the action $I_s$. Here, $m_{\rm eff}$ and $V_{\rm eff}$ are constant effective parameters, and the new conjugate pair of coordinates are $\Theta \in [-\pi, \pi)$ and $P = I-I_s$ \cite{sup1}. Thus, in the frame moving along the resonant orbit, a particle behaves like an electron in a spatially periodic potential with $s$ sites. 
Quantizing the Hamiltonian $H_i$, we obtain its eigenstates in the form of Bloch waves and for $s \gg 1$ the corresponding eigenenergies form energy bands. Focusing on the first energy band, Wannier states $w_{i,\alpha}(\Theta)$ can be defined; they are localized \cite{MarzariEtAl2012} in individual sites ($\alpha = 1,\dots,s$) of the effective potential $- V_{\rm eff} \cos (s\Theta)$. 
In the original variables, these Wannier states are localized wave packets $w_{i,\alpha}(\theta-\omega t/s)$ moving along the resonant orbit. On a detector located close to the resonant orbit (i.e.\ at fixed $\theta$ and $I\approx I_s$), the probability of detection of a particle prepared in an eigenstate of $H_i$, e.g. $e^{i{\cal K}(\theta-\omega t/s)}\sum_\alpha w_{i,\alpha}(\theta-\omega t/s)$, changes periodically in time and reflects a crystalline structure in the time domain --- $\cal K$ can be interpreted as the time-quasimomentum. 

\begin{figure}
\includegraphics[width=84mm]{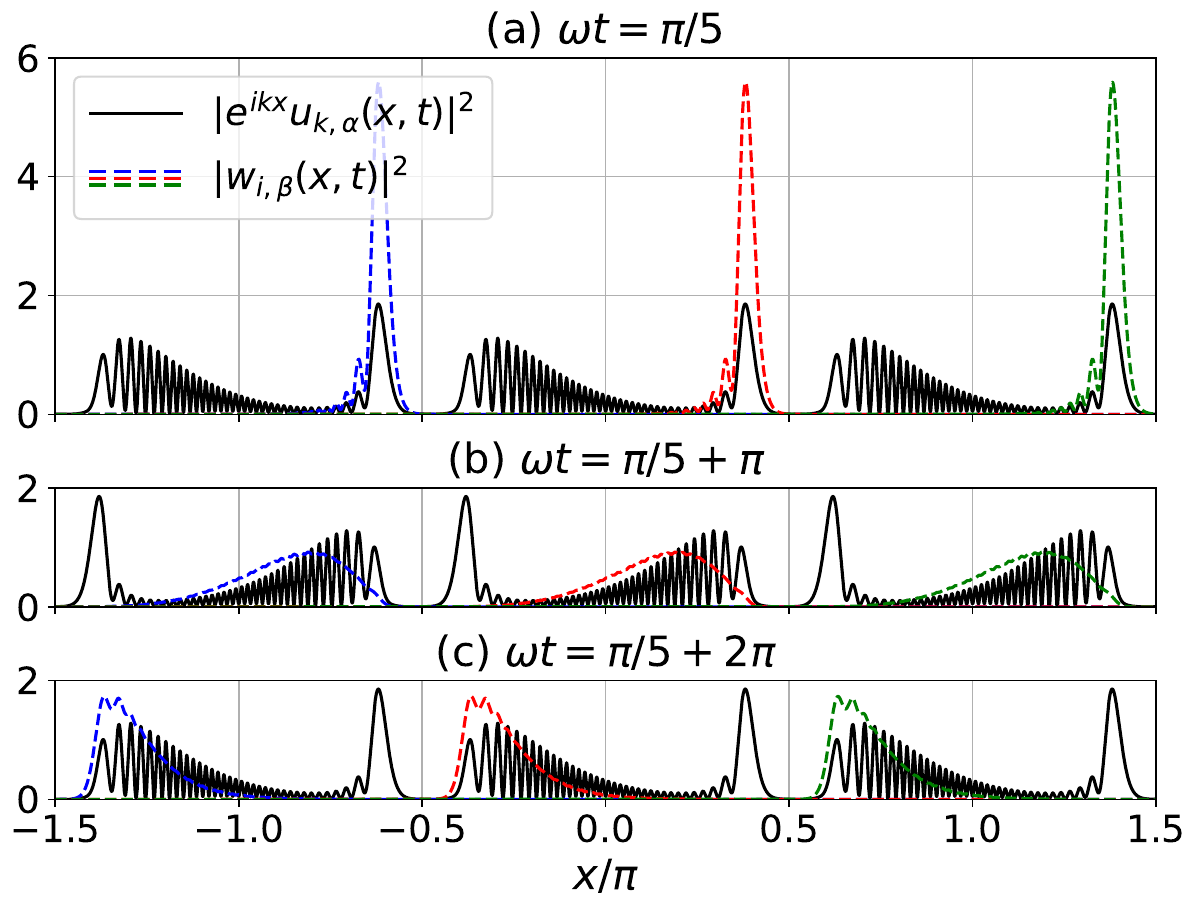}
\caption{Probability densities of the $k = 0$ Floquet-Bloch state $|u_{0,\alpha}(x,t)|^2$ (black line) and the individual Wannier states $|w_{i,\beta}(x,t)|^2$ (dashed colored lines) for a particle in the resonantly vibrating optical lattice potential (\ref{H_basic}), at different moments of time, i.e. at (a) $\omega t = \pi/5$, (b) $\omega t = \pi/5+\pi$ and (c) $\omega t = \pi/5+2\pi$. The parameters of the system are the following: $s = 3$, $V_0 = 4320$, $\omega = 240$ and $\lambda = 0.01$. Three sites of the optical lattice potential are shown only and in each site only one Wannier state (out of three for $s=3$) is presented. Note that the Floquet-Bloch states are periodic with the period $\omega T = 2\pi$, whereas the period of the Wannier states is three times longer.}
\label{one}   
\end{figure}

\begin{figure}
\includegraphics[width=84mm]{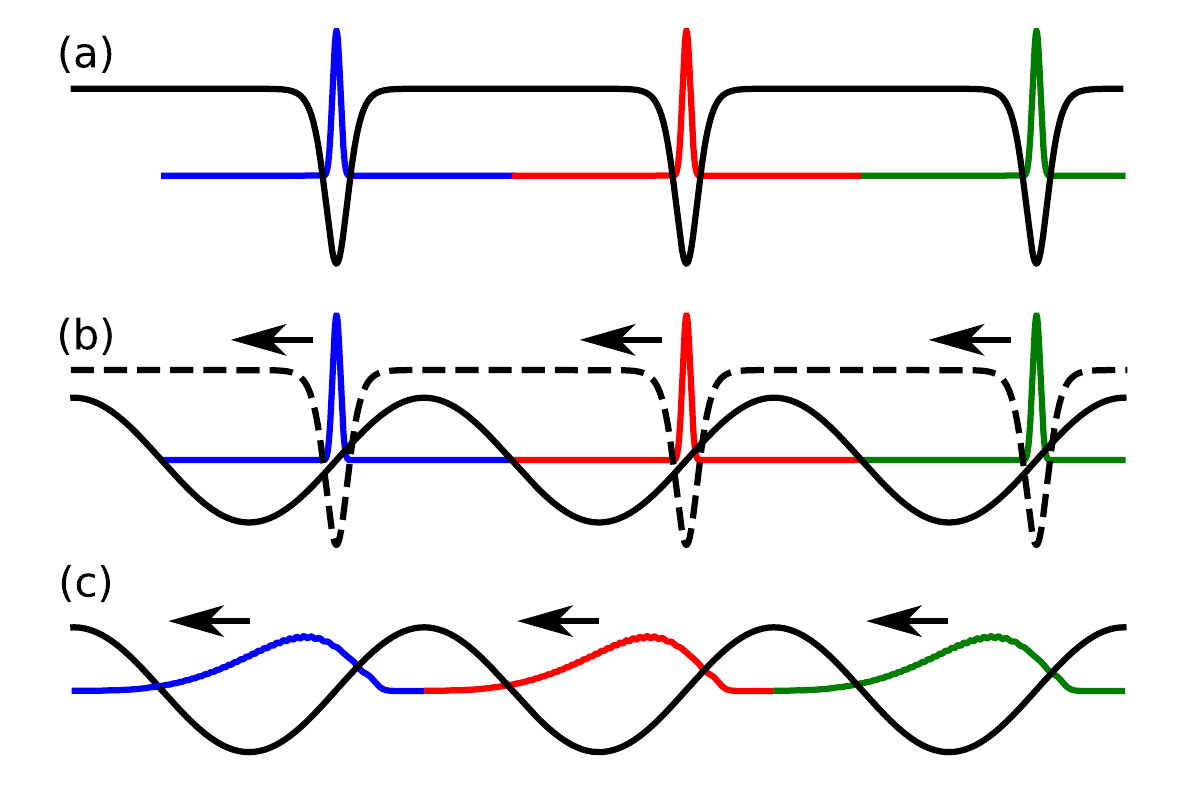}
\caption{Loading ultracold atoms to the resonant manifold: (a) The atoms are first loaded into the ground state of an auxiliary superlattice that consists of a periodic sequence of narrow potential wells --- their widths are chosen such that the lowest-band Wannier functions (shown by the blue, red and green full lines) approximate the targeted Wannier wave packets $w_{i,\alpha} (x)$. (b) Next, the auxiliary lattice (dashed black line) is turned off and the targeted shaken lattice (solid black line) is turned on to place the prepared Wannier states (red lines) at the classical turning points; the profiles shown here correspond to $\omega t = 0$.  (c) Wannier states undergo periodic oscillations; the profiles shown here correspond to $\omega t = \pi/5+\pi$, cf.\ Fig.~\ref{one}~(b).} \label {two}   
\end{figure}

The easiest way to describe the entire system beyond a single site is to apply a quantum version of the secular approximation \cite{Berman1977}. Due to the periodicity of the system in space and in time we may look for time-periodic Floquet states (i.e.\ eigenstates of the Floquet Hamiltonian ${\cal H} = H - i\partial_t$ \cite{Shirley1965,Buchleitner2002}) that describe resonant motion of a particle and are Bloch waves of the form $e^{ikx} u_{k,\alpha}(x,t)$ where $k$ is the usual quasimomentum while $\alpha = 1, \dots,s$ labels resonant Floquet states. The wave functions $u_{k,\alpha}(x,t+2\pi/\omega) = u_{k,\alpha}(x,t) = u_{k,\alpha}(x+\pi,t)$ fulfill the eigenvalue problem
\be
  \left[ H^{(0)} + (p_x+k)\lambda\omega\sin\omega t - i\partial_t \right] u_{k,\alpha} = E_{k,\alpha}u_{k,\alpha},
\ee
where the unperturbed part of the Hamiltonian reads $H^{(0)}=(p_x+k)^2+V_0\sin^2x$ and $E_{k,\alpha}$ are quasienergies \cite{Buchleitner2002}.
We choose the eigenstates $\psi_{k,n}(x)$ of the unperturbed Hamiltonian, $H^{(0)}\psi_{k,n}=E_{k,n}^{(0)}\psi_{k,n}$, as the basis for the Hilbert space, perform the time-dependent unitary transformation $\psi_{k,n}' = e^{-in\omega t/s}\psi_{k,n}$ (which is the quantum analog of the classical transformation to the frame moving along the resonant orbit) and finally neglect all time-oscillating terms \cite{sup1}. This yields the following matrix elements of the effective Floquet Hamiltonian
\begin{equation}
\begin{split}
\label{matrixH_F}
  \la \psi'_{k,n'} |{\cal H}|\psi'_{k,n} \ra &\approx \left( E_{k,n}^{(0)} - n\frac{\omega}{s}\right) \delta_{n'n} 
  - \la \psi_{k,n'}|p_x|\psi_{k,n}\ra \\
  &\times \frac{i\lambda\omega}{2} \left(\delta_{n',n-s}-\delta_{n',n+s}\right).
\end{split}
\end{equation}
Diagonalization of (\ref{matrixH_F}) allows us to obtain the resonant Floquet-Bloch states $e^{ikx}u_{k,\alpha}(x,t)$.
Proper superpositions of $e^{ikx}u_{k,\alpha} (x,t)$ form time-periodic Wannier states $w_{i,\alpha}(x,t)$ which are localized wave packets moving along the resonant orbit with the period $s2\pi/\omega$. To demonstrate the temporal and spatial behavior of these states, in Fig.~\ref{one} we plot the probability densities of the $k = 0$ Floquet-Bloch state $|u_{0,\alpha}(x,t)|^2$ and the three Wannier states. We note that: (i) the probability densities $|u_{k,\alpha}|^2$ change only a little with $k$, and (ii) in order to obtain the appropriate form of the Floquet-Bloch states in the lab frame one needs to apply the unitary transformation $e^{-ip_x \lambda\cos\omega t}$. However, we consider very small amplitudes, e.g.\ $\lambda = 0.01$, thus the actual changes imposed by this transformation are negligible. In the subspace spanned by the Wannier states $w_{i,\alpha}(x,t)$,  
the Floquet Hamiltonian
takes the form of the tight-binding model \cite{sup1}
\be
  \hat{\mathcal{H}} = -\frac12 \sum_{i,\alpha,j,\beta} J^{j,\beta}_{i,\alpha}  \hat{a}^\dag_{j,\beta} \hat{a}_{i,\alpha},
\label{tb}
\ee
where $\hat{a}^{(\dag)}_{i,\alpha}$ is the annihilation (creation) operator acting on the site $(i,\alpha)$.
The tunneling amplitudes $J^{j,\beta}_{i,\alpha}=-(2/sT)\int_0^{sT}dt\la w_{j,\beta}|{\cal H}|w_{i,\alpha}\ra$ 
with $T = 2\pi/\omega$
describe hopping transitions between sites of the optical lattice (Latin labels) and between the time-lattice sites (Greek labels) and are dominant for nearest-neighbor hopping (note that the coefficients $J_{i,\alpha}^{i,\alpha}=\rm const.$ do not depend on either $i$ or $\alpha$ and are the on-site energies). 
The range of the validity of the quantum secular Hamiltonian (\ref{matrixH_F}) can be examined by checking if the classical secular Hamiltonian reproduces quantitatively the exact classical motion \cite{sup1}. 
If it is so, the quantum counterpart is also valid. 
This implies that we are interested in the regime where eigenenergies of $H^{(0)}$ support multiple bands, which is provided by large lattice depths $V_0 \simeq 10^3$. 
Experimentally, this regime is attainable but in order to deal with appreciable hopping between sites of the optical lattice, the resonant condition for time-periodic shaking of the lattice must correspond to a highly excited band with energy $E\lesssim V_0$ and this is the regime we explore here. Note that we are interested in a resonant coupling between highly excited bands of an optical lattice not in a resonant coupling between the lowest bands as, e.g., in \cite{Sowinski2012,Lacki2013,Li2016}. 
In the numerical example (see below) we have $|J_{i,\alpha}^{i+1,\alpha}| \sim 10^{-3}$ which is about an order of magnitude larger than the incoherent scattering rate for, e.g., $^{87}$Rb atoms in the presence of the CO$_2$ laser radiation at wavelength $10.6\,\mathrm{\mu m}$
\cite{co2laser}.
In order to load ultracold atoms to the resonant manifold, the atomic cloud must be initially prepared in an auxiliary static optical lattice consisting of narrow wells so that the width of the corresponding Wannier states matches the width of the Wannier wave packets $w_{i,\alpha}$. Next the auxiliary lattice should be turned off and the vibrating optical lattice (slightly displaced with respect to the auxiliary lattice) should be turned on. This results in a state where for each $i$ only one site of the time lattice is occupied, see Fig.~\ref{two}. The appropriateness and efficiency of a similar loading process was explored in Refs.~\cite{Giergiel2018a,Kuros2020}.

{\it 6D time-space crystals.}---Generalization of 2D \tsc\ to 4D or 6D \tsc\ is straightforward. Indeed, let us consider a 3D optical lattice vibrating along all three orthogonal directions. Then, the Hamiltonian of a particle in the frame vibrating with the lattice is the sum $H(x,p_x,t)+H(y,p_y,t)+H(z,p_z,t)$ where $H$ is given in (\ref{H_basic}). Because there is no coupling between different degrees of freedom of a particle, in order to describe 3D resonant motion of a particle one may use the results obtained in the case of the 2D \tsc. 
We define the Wannier states $W_{\vec i,\vec \alpha}(\vec r,t) = w_{i_x,\alpha_x}(x,t) w_{i_y,\alpha_y}(y,t) w_{i_z,\alpha_z}(z,t)$, where $w_{i,\alpha}$ are Wannier wave packets constructed in the previous paragraph, and derive the 6D tight-binding model of the same form as (\ref{tb}) but with the indices $i$ and $\alpha$ generalized to three-component vectors $\vec i=(i_x,i_y,i_z)$ and $\vec \alpha=(\alpha_x,\alpha_y,\alpha_z)$, see \cite{sup1}. It means that in each site of the 3D optical lattice we realize the $s\times s\times s$ net of periodically moving wave packets $W_{\vec i,\vec \alpha}(\vec r,t)$ which is associated with a 3D temporal lattice. The entire system of all such sites of the 3D optical lattice forms a 6D crystalline structure. In the following, we show how 6D \qhe\ can be realized in such a 6D \tsc\ by first setting the stage with the realization of synthetic gauge fields that support the 2D \qhe\ in a 2D \tsc.

{\it Quantum Hall effect in 2D time-space crystals.}---In ultracold atoms prepared in a 2D time-independent optical lattice, the \qhe\ was realized with the help of the photon-assisted tunneling against a potential gradient created along one of the two spatial directions \cite{AidelsburgerEtAl2013,miyake13harper}, alternatively -- along a synthetic dimension \cite{CeliEtAl2014} encoded by the atom's internal state \cite{StuhlEtAl2015,ManciniEtAl2015}. Importantly, the hopping amplitude acquired a phase that could be controlled by changing the angle between the laser beams. It allowed one to realize a system where tunneling of an atom around an elementary plaquette resulted in a Aharonov-Bohm phase \cite{Goldman2014} that defines an artificial flux. The energy bands of such a system are characterized by the first Chern number that determines the quantization of the Hall conductivity. Here we show that the concept of photon-assisted tunneling against a potential gradient can be extended to \tsc, however, the potential tilt is now realized along the temporal direction.

We thus consider an atom in the vibrating optical lattice potential of Eq.~(\ref{H_basic}) in the presence of an additional secondary optical lattice vibrating in time in the same way. Also, the previously considered harmonic vibration must now be modulated according to
a tailored signal $f_x(t) = \sum_{n\ne 0} f_n^{(x)} e^{in\omega t/s}$. The total Hamiltonian of the system reads 
\be
  \bar H(x,p_x,t) = H(x,p_x,t) + V_1\sin^2(2x+\phi) + p_x f_x(t),
\label{second}
\ee
where $H$ is given in (\ref{H_basic}), $V_1\ll V_0$ and $\phi\ne 0$. The goal is to engineer the effective Hamiltonian with an additional linear term, $\bar H_i = P^2/(2\bar m_{\rm eff}) - \bar V_{\rm eff} \cos(s\Theta) + U_x\Theta$, where
\be
  U_x\Theta = \sum_{n\ne 0}f_{-n}^{(x)}p_n(E)e^{in\Theta},
\label{linearV}
\ee
and $U_x 2\pi/s$ is smaller than the energy gap between the first and second energy bands of $\bar H_i$ with $f_x(t) = 0$. This requires that the additional modulation $f_x(t)$ of the lattices' motion is chosen in such a way that $f_{-n}^{(x)}p_n(E)$ are the Fourier components of the linear potential $U_x \Theta$. Here $p_n(E)$ are Fourier components of $p_x(t)$ corresponding to the motion of a particle on the unperturbed resonant orbit, i.e. $p_x(t) = \sum_n p_n(E) e^{in\Omega(E)t}$ with $\Omega(E) = \omega/s$ \cite{sup1}. Note that for $V_1 = 0$, the even components $p_{2m}(E) = 0$, therefore in order to realize the linear potential (\ref{linearV}) we have to add the secondary optical lattice, cf.~(\ref{second}), to provide $p_{2m}(E)\ne 0$. 
The presence of the linear potential in the effective Hamiltonian $\bar H_i$ describing a single site indicates that in the frame
moving along the resonant orbit a particle experiences a potential tilt. 
Eigenstates of the quantized version of $\bar H_i$ are localized in $\alpha = 1,\dots,s$ sites of the potential $-\bar V_{\rm eff}\cos(s\Theta)$; and they form a localized basis similar 
to the Wannier state basis which we also denote $w_{i,\alpha}(\Theta)$ but hopping between sites with different $\alpha$ is suppressed.

The quantized version of the classical secular Hamiltonian $\bar H_i$ allows us to find how to shake the optical lattices in order to turn off hopping of a particle between Wannier states $w_{i,\alpha}$ and $w_{i,\beta}$. The predictions based on the classical secular approach are confirmed by the results of the quantum secular approximation which yields the tight-binding Hamiltonian of the form of (\ref{tb}) where $J_{i,\alpha}^{i,\beta}=0$ for $\beta\ne\alpha$ and $J_{i,\alpha}^{i,\alpha}\approx \alpha2\pi U_x/s$. It means we are able to suppress hopping between time-lattice sites by tilting the \tsc\ along the temporal direction. 

The last stage of the realization of the \qhe\ in the 2D \tsc\ is to reestablish the suppressed hopping but with complex tunneling amplitudes  $J_{i,\alpha}^{i,\beta}\propto e^{i\varphi_i}$ where the phase $\varphi_i$ is a linear function of the site index $i$. To this end we apply two laser beams with slightly different frequencies (i.e. the difference of the photon energies matches the difference of the on-site energies,  $2\pi U_x/s$, of neighboring sites of the tilted time-lattice) propagating at an angle to the $x$ axis so that the difference of the wave vectors points along the $x$ direction and its length is denoted by $\Delta k_{\rm ph}$. 
Absorption of a photon by an atom from one beam and emission of a photon into the other beam is described by the spatially dependent Rabi frequency $\Omega_{\rm ph}e^{i\Delta k_{\rm ph}x}$. Following a well established approach (see e.g.\ Refs.\  \cite{Jaksch2003,Gerbier2010,kolovsky11,AidelsburgerEtAl2013,miyake13harper,Eckardt16review} and in particular the detailed account in Sec.~8.4.3 of Ref.~\cite{Goldman2014}), we find that the presence of such a two-photon process modifies the parameters of the tight-binding model (\ref{tb}), i.e. now we have $J_{i,\alpha}^{i,\alpha}\approx \rm constant$ and non-zero
\be
\label{eq:rabi}
  J_{i,\alpha}^{i,\alpha+1}=\int_0^{sT}\frac{dt}{sT}\int dx \;w^*_{i,\alpha+1}(x,t)\Omega_{\rm ph}e^{i\Delta k_{\rm ph}x}w_{i,\alpha}(x,t).
\ee
Note that $J_{i,s}^{i,1} = 0$ because the photon-assisted tunneling is not able to compensate the difference between the on-site energies of the $\alpha=1$ and $\alpha=s$ sites.
Thus, we deal with the ribbon geometry characterized by open boundary conditions along the time direction. Importantly the complex phase of $J_{i,\alpha}^{i,\alpha+1}$ changes with $i$ and it can be controlled by changing $\Delta k_{\rm ph}$. Thus, an atom acquires a phase when it tunnels around a plaquette of the \tsc. Such a phase can be associated with a magnetic-like flux and the system can reveal the 2D quantum Hall effect. That is, in the limit of $s\rightarrow\infty$, the tight-binding model describes a bulk system and depending on the value of the magnetic-like flux different number of energy bands form which can be characterized by the non-vanishing first Chern number $\nu_1^{(x)}=(2\pi)^{-1}\int_{\rm BZ}dkd{\cal K}\Omega^{xt}$ which is the integral of the Berry curvature, $\Omega^{xt}=\partial_k{\cal A}_{\cal K}-\partial_{\cal K}{\cal A}_k$, over the Brillouin zone (BZ) where ${\cal A}_{k}=\la n_{k,{\cal K}}|\partial_{k}|n_{k,{\cal K}}\ra$ and ${\cal A}_{{\cal K}}=\la n_{k,{\cal K}}|\partial_{{\cal K}}|n_{k,{\cal K}}\ra$ define the Berry connection and $e^{i(kj+{\cal K}\alpha)}|n_{k,{\cal K}}(j,\alpha)\ra$ is a Bloch wave belonging to the $n$-th band \cite{ThoulessEtAl1982,LeeEtAl2018,PetridesEtAl2018}.

We would like to stress that the topological \tsc\ considered here is described by a 2D tight-binding model and is distinct from a system that realizes a 1D topological pump \cite{Thouless1983,Lohse2016pump,LohseEtAl2018,PetridesEtAl2018} based on a 1D model whose parameter changes periodically in time. In other words, the developed model provides a versatile platform for studies of condensed matter phases of 2D crystalline structures, supported by Hubbard-type models \cite{sup1,Dutta2015}.

\begin{figure} 	            
\includegraphics[width=84mm]{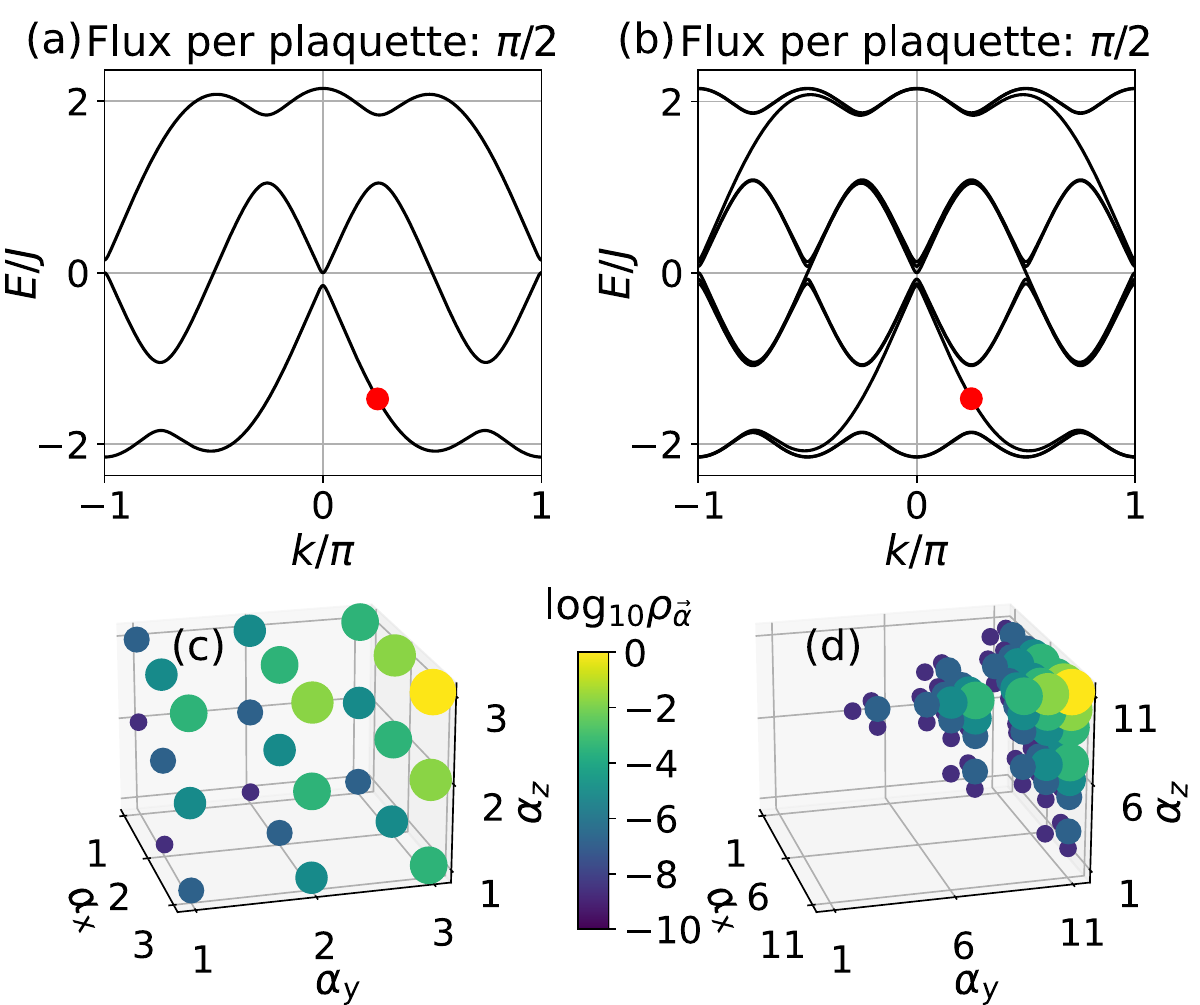}
\caption{Energy dispersion obtained from a 2D tight-binding model of a three-leg [panel (a)] and $11$-leg [panel (b)] ribbon with hopping parameters $J=|J_{i,\alpha}^{i+1,\alpha}|$ and $|J_{i,\alpha}^{i,\alpha+1}|= 0.4 J$ in the long and short directions, respectively, and $\pi/2$ flux per plaquette encoded by Peierls phases \cite{CeliEtAl2014}. Formation of a topological subband structure with edge modes crossing the gaps is apparent. Combining three such 2D models into a 6D model, its edge mode is built as a product of states at quasimomentum $k_x = k_y = k_z = \pi/4$ marked by the red dot in (a) and (b). The probability density $\rho_{\vec \alpha}$ of this mode in a three-leg [panel (c)] and $11$-leg [panel (d)] ribbon is plotted as a function of the time-lattice indices $\vec \alpha=(\alpha_x,\alpha_y,\alpha_z)$ and illustrates its localized nature with respect to all three directions, i.e.\ on the edge of the 6D ribbon. Parameters that can realize such a ribbon are $\omega = 236.2$, $V_0 = 4320$, $V_1 = 0.1 V_0$, $\phi = \pi/8$ and $U_x = 1$. The first ten harmonics for the tilted potential are included, cf.~(\ref{linearV}). This gives the hopping parameter $J \sim 10^{-3}$.
\label{three}}
\end{figure} 

{\it Quantum Hall effect in 6D time-space crystals.}---The idea for the realization of the topological 2D \tsc\ can be generalized to higher dimensions. Instead of a 1D optical lattice we can consider a particle in a 3D lattice vibrating in three orthogonal directions which is described by the Hamiltonian $\bar H(x,p_x,t)+\bar H(y,p_y,t)+\bar H(z,p_z,t)$ where $\bar H$ is given in (\ref{second}). 
For each degree of freedom of a particle one can apply the same idea of the photon-assisted tunneling as in the case of the 2D \tsc\ and realize a 6D counterpart of the tight-binding model (\ref{tb}) where the complex phase $\varphi_{\vec i}$  of the tunneling amplitudes $J_{\vec i,\vec \alpha}^{\vec i,\vec \beta}\propto e^{i\varphi_{\vec i}}$ changes with a change of the optical lattice indices $\vec i$. 
Note that, if $f_{x,y,z}(t)$ have different amplitudes, the resulting potential tilts $U_{x,y,z}$ are different and each pair of three pairs of the laser beams can be tuned to induce hopping along one spatial direction only. The 6D quantum Hall effect can be observed if the third Chern numbers $\nu_3$ of energy bands of such a tight-binding model do not vanish. The spatial degrees of freedom of a particle are decoupled which implies that eigenstates of the 6D tight-binding model are products of 2D Bloch waves, i.e. $e^{i(\vec k\cdot \vec j+\vec {\cal K} \cdot\vec \alpha)}|n_{k_x,{\cal K}_x}(j_x,\alpha_x)\ra|n_{k_y,{\cal K}_y}(j_y,\alpha_y)\ra|n_{k_z,{\cal K}_z}(j_z,\alpha_z)\ra$, and consequently the third Chern number is a product of the first Chern numbers $\nu_3=\nu_1^{(x)}\nu_1^{(y)}\nu_1^{(z)}$. The third Chern number determines how the third order current response depends on an external electromagnetic perturbation \cite{PetridesEtAl2018}. The topological character of the 6D \tsc\ can be illustrated by the presence of topologically protected edge states. For a finite resonance number $s$, the photon-assisted tunneling reestablishes hopping between time lattice sites but not between the first $\alpha_x,\alpha_y,\alpha_z=1$ and the last $\alpha_x,\alpha_y,\alpha_z=s$ sites which means we deal with the open boundary conditions and edges in the \tsc.

In Fig.~\ref{three} we plot the energy dispersions for a narrow ($s = 3$) and broader ($s = 11$) 2D ribbon with open boundary conditions on the $\alpha = 1$ and $\alpha = s$ edges. In the broad ribbon, the formation of topological band structure is evidenced by the presence of chiral edge states that cross the gap. The number of such left and right propagating modes is equal to the sum of the Chern numbers characterizing the bands below the gap. The band structure of the narrow ribbon clearly shows precursors of these edge modes. While the chiral nature of such edge modes has been demonstrated in the literature, let us focus on their edge-like nature in the synthetic temporal dimensions. We thus define the standard on-site densities 
$\rho_{\vec{i},\vec{\alpha}} = \la\psi|\hat{a}^{\dag}_{\vec{i},\vec{\alpha}} \hat{a}_{\vec{i},\vec{\alpha}}|\psi\ra$ and their projections to the synthetic dimensions $\rho_{\vec{\alpha}} = \sum_{\vec{i}} \rho_{\vec{i},\vec{\alpha}}$. The distributions of these densities, for eigenstates $|\psi\ra$ corresponding to eigenenergies in energy gaps, are shown in panels (c) and (d) of Fig.~\ref{three} and demonstrate the formation of edge states in the time dimension
\footnote{We note that in narrow ribbons (e.g. with $s = 3$) the interplay of the harmonic effective potential with the linear tilt may introduce some nonuniformity in the tunneling amplitudes and fluxes along the short direction. Such imperfections can be corrected by periodically modulating the Rabi frequency in Eq.~(\ref{eq:rabi}).}. 
Such edge states can be prepared experimentally by means of the method illustrated in Fig.~\ref{two}.
 
To conclude, we have shown that by combing time and space crystalline structures it is possible to realize 6D time-space crystals. Six-dimensional condensed matter physics is attainable if a 3D spatially periodic system is resonantly driven in time. As an example we have described a route for the realization of gauge fields and observation of the 6D quantum Hall effect. Is 6D time-space electronics around the corner?

G.Z.\ and C.-h.F.\ made an equal contribution to the letter. We thank Daria Cegie\l{}ka for the collaboration at the early stage of the project and Tomasz Kawalec and Peter Hannaford for the helpful discussions. 
Support of the National Science Centre, Poland via Project No.~2018/31/B/ST2/00349 (C.-h.F.) and QuantERA programme No.~2017/25/Z/ST2/03027 (K.S.) is acknowledged. 
The work of G.Z.\ and E.A.\ was supported by the European Social Fund under Grant No.\ 09.3.3-LMT-K-712-01-0051. 




\newpage

\centerline{\bf \LARGE Supplemental Material}

\vspace{1.cm}

In this Supplemental Material, we: 
\bi
\item 
introduce the basic Hamiltonian used in the letter,
\item
present the classical and quantum versions of the secular approximation for the resonant dynamics of a particle in a periodically shaken optical lattice potential,
\item 
show how to derive the tight-binding model for the time-space crystalline structure, 
\item
explain in detail how to generalize a two-dimensional (2D) crystalline structure to a 6D one.
\ei
However, we begin with a description of the concept of condensed matter physics in time crystals \cite{sSacha2017rev,sGuo2020,sSachaTC2020}.

\section{Condensed matter physics in time crystals}

Discrete time crystals are periodically driven quantum many-body systems which are able to break spontaneously discrete time translation symmetry of the drive \cite{sSacha2017rev,sSachaTC2020}. It is a temporal demonstration of one of the important properties of solid state systems. It turns out that apart from the discrete time crystals and the spontaneous symmetry breaking, it is also possible to realize other condensed matter phenomena in the time dimension. In ordinary space crystals we deal with periodic distributions of atoms in space which can be detected at a fixed moment of time. When we switch from space to time crystals, the roles of space and time have to be interchanged. We fix a position in space (i.e. we choose a space point where we locate a detector) and ask if a detector clicks periodically in time. We are not interested in trivial time evolution of a system but in time periodic behavior which can be described by temporal counterparts of solid state models. It has been already shown that Anderson localization, many-body localization, quasi-crystal behavior or topological phases can be observed in the time dimension \cite{sSacha2017rev,sGuo2020,sSachaTC2020}. It is even possible to realize time lattices with properties of two- or three-dimensional space crystalline structures \cite{sGiergiel2018}. There, different locations of a detector allows one to probe multi-dimensional dynamical lattices along different directions. We would like to stress that crystalline structures we consider here are not emergent phenomena but they are created by a proper periodic driving of systems in time \cite{sSacha2017rev,sGuo2020,sSachaTC2020}. It is a similar situation like in photonic crystals which do not emerge spontaneously because periodic modulation of the refractive index in space is imposed externally \cite{sJoannopoulos_Book}.

In the present letter we show how to combine together crystalline structures in space and in time and create even 6D time-space crystals. As an example of non-trivial 6D physics we have chosen a 6D quantum Hall effect but the single-particle and many-body condensed matter phenomena demonstrated already in time crystals \cite{sSacha2017rev,sGuo2020,sSachaTC2020} can be also realized in 6D time-space crystalline structures. 

\section{Hamiltonian for a particle in a periodically shaken optical lattice potential}

Let us consider an atom in a periodically shaken optical lattice potential described by the scaled dimensionless Hamiltonian 
\be
  H(x,p_x,t) = p_x^2 + V_0 \sin^2 (x - \lambda \cos\omega t).
\ee
Here we choose to work in the recoil units for the energy $\hbar^2 k_L^2 / 2m$ and length $1/k_L$, with $k_L$ being the wave number of laser beams that create the optical lattice. The depth of the lattice is $V_0$ and its position is modulated periodically in time with the amplitude $\lambda$ and frequency $\omega$. It is convenient to switch to the frame moving with the lattice; in the classical case this is done by means of the time-dependent canonical transformation $x' = x - \lambda\cos\omega t$ and $p_x' = p_x$ while in the quantum case using the unitary transformation $e^{ip_x\lambda \cos\omega t}$. This leads to the Hamiltonian 
\be
  H(x,p_x,t) = p_x^2 + V_0 \sin^2x + p_x \lambda \omega \sin\omega t,
\label{SH_basic}
\ee
where we drop the primes and use the same symbols $x$ and $p_x$ for the transformed position and momentum variables. The Hamiltonian (\ref{SH_basic}) is the starting point for the entire analysis performed in the letter.

We note that by means of a further transformation the momentum shift can be traded for a term $-x\lambda\omega^2\cos\omega t$ describing a homogeneous inertial force \cite{sArimondoEtAl2012}. However, we choose not to follow this course in order to preserve the explicit spatial translational symmetry.

\section{Classical secular approximation approach}

For the sake of reference, let us start with the unperturbed situation, i.e., a stationary lattice ($\lambda = 0$) and a classical particle of energy $E < V_0$ undergoing periodic oscillations in the vicinity of a single potential minimum with an energy-dependent frequency $\Omega(E)$. To simplify the classical description it is convenient to perform the canonical transformation to the action-angle variables \cite{sLichtenberg1992}, 
\begin{subequations}
\begin{align}
  I &= \frac{1}{2\pi} \oint p_x (E,x) dx, \\
  \theta &= \frac{\partial}{\partial I} \int \limits_{x_0}^x p_x(E,x') dx',
\end{align}
\end{subequations}
with $x_0$ set to the left classical turning point.
Then the unperturbed Hamiltonian depends on the new momentum $I$ (the action) alone, i.e. $H_{0} = p_x^2 + V_0\sin^2 x = H_{0}(I)$, and the  solution of the Hamilton equations of motion is trivial: $I = \mathrm{const}.$, while the new coordinate (the angle) is changing at a constant rate $\theta(t) = \Omega(E)t + \theta(0)$ with $\Omega(E(I)) = dH_{0}(I)/dI$. 

Now let us turn on the shaking, which couples to the momentum $p_x (\theta, I)$, and assume that the frequency $\omega$ fulfills the resonant condition $\omega = s\,\Omega(E(I_s))$ where $s$ is an integer and $I_s$ is the resonant value of the action. Such a $s:1$ resonant driving of a particle can be accurately described by the secular approximation provided the time-periodic perturbation is weak \cite{sLichtenberg1992,sBuchleitner2002,sSacha2017rev,sSachaTC2020}. In order to obtain the effective Hamiltonian that describes the motion of a particle in the vicinity of the resonant orbit, we first switch to the frame moving along the resonant orbit, $\Theta = \theta-\omega t/s$, which results in 
\be
\begin{split}
  H(\Theta,I,t) &= H_0(I) - \frac{\omega I}{s} \\
  &+ p_x(\Theta+\omega t/s,I)\lambda\omega \sin\omega t.
\end{split}
\ee
As the momentum of a particle on the resonant orbit is periodic with respect to $\theta$, we perform the Fourier expansion,
\be
  p_x (\Theta + \omega t/s,I) = \sum_n p_n(I) e^{in(\Theta + \omega t/s)},
\ee
and finally average the Hamiltonian over the time keeping $\Theta$ and $I$ fixed because they are slowly varying variables if we stay close to the the resonant orbit, i.e. for $P = I-I_s\approx 0$ \cite{sLichtenberg1992,sBuchleitner2002,sSacha2017rev,sSachaTC2020}. This yields the effective time-independent Hamiltonian
\be
  H_i = \left[H_0(I_s) - \frac{\omega I_s}{s}\right]
  + \frac{P^2}{2m_{\rm eff}} - V_{\rm eff}\cos(s\Theta),
\label{Sclass_sec}
\ee
where $\Theta \in [-\pi,\pi)$, $V_{\rm eff} = \lambda \omega \, |p_s(I_s)|$ and we have performed the Taylor expansion for $I$ around $I_s$ up to the second order, which allowed us to define the effective mass $m_{\rm eff}^{-1} = d^2H_0(I_s)/dI_s^2$. The constant term in the square brackets in (\ref{Sclass_sec}) can be omitted because it does not influence the dynamics.
The Hamiltonian (\ref{Sclass_sec}) describes the resonant motion of a particle in the $i$-th site of the shaken optical lattice. Its form shows that for $s\gg 1$, a particle behaves like an electron moving in a periodic potential created by ions in a solid state crystal. The Hamiltonian (\ref{Sclass_sec}) describes a crystalline structure but in the moving  frame. When we return to the laboratory frame, a condensed matter-like behavior observed versus $\Theta$ in the moving frame will be observed in the time domain. Indeed, the transformation to the moving frame is linear in time, i.e. $\Theta=\theta-\omega t/s$. Thus, if we fix the position in the laboratory frame (i.e. we locate a detector at fixed $\theta$ close to the resonant orbit), then the probability of clicking of the detector versus time will reflect the same behavior as the probability for detection a particle as a function of $\Theta$ in the moving frame.
This situation can be interpreted as engineering of a synthetic dimension, related to the time variable rather than atom's internal degrees of freedom \cite{sStuhlEtAl2015,sManciniEtAl2015,sCeliEtAl2014}. Note that for a particle moving in the sinusoidal potential $V_0\sin^2x$ even Fourier components of the momentum vanish, $p_{2n}(E)=0$. Consequently only odd numbers $s$ of sites in the periodic potential in (\ref{Sclass_sec}) can be realized.

In the letter we are also interested in the realization of a tilted periodic effective potential, i.e. when the effective Hamiltonian describing resonant motion of a particle in the $i$-th lattice has the form
\be
  \bar H_i = \frac{P^2}{2\bar m_{\rm eff}} - \bar V_{\rm eff}\cos(s\Theta) + U_x\Theta.
\label{Sclass_sec_bar}
\ee
It can be done if one adds a weak secondary optical lattice and starts shaking the entire potential in an appropriate manner in time. With this goal in mind, the Hamiltonian (\ref{SH_basic}) is modified to read
\be
\begin{split}
  \bar H(x,p_x,t) &= p_x^2+V_0\sin^2x+V_1\sin^2(2x+\phi) \\
  &+ p_x \left[\lambda\omega\sin\omega t + f_x(t)\right],
\label{SH_basic_bar}
\end{split}
\ee
with  $V_1\ll V_0$ and
\be
  f_x(t) = \sum_{n\ne 0}f_n^{(x)}e^{in\omega t/s}
\ee
describing a weak modulation of the harmonic vibration of the entire lattice. Introducing the action-angle variables corresponding to the new unperturbed Hamiltonian $\bar H_0 = p_x^2 + V_0\sin^2 x + V_1 \sin^2(2x+\phi)$ we again rely on the same secular approximation and obtain the effective Hamiltonian (\ref{Sclass_sec_bar}) where 
\be
  U_x \Theta = 
  U_x\sum_{n\ne 0}\frac{i(-1)^n}{n}e^{in\Theta}
  = \sum_{n\ne 0}f_{-n}^{(x)}p_n(E)e^{in\Theta}.
\label{SlinearV}
\ee
That is, $f_x(t)$ is chosen so that $f_{-n}^{(x)}p_n(E)$ are Fourier components of the linear potential. Note, that if $p_n(E)\ne 0$ for all $n$, it is always possible to adjust $f_{-n}^{(x)}$ so that the products $f_{-n}^{(x)}p_n(E)$ take values we need. The role of the secondary lattice is now clear because for $V_1 = 0$ we would not be able to reproduce the linear potential in (\ref{SlinearV}) due to the fact that the even components $p_{2n}(E) = 0$. When the secondary lattice is on, all components $p_n$ are non-zero and we can engineer any effective potential, in particular the linear one. Note that despite the fact that the function $f_x(t)$ appears in the original Hamiltonian (\ref{SH_basic_bar}) as a linear term, it can result in very different effective potentials depending how we choose the Fourier components $f_n^{(x)}$.

We can locate a particle in different local minima of the effective potential in Eq.~(\ref{Sclass_sec_bar}), i.e. at $\Theta\approx j2\pi/s$ where  $j=1,\dots,s$ . In the laboratory frame when we fix position of a detector close the resonant orbit, different locations correspond to different delays in time, $j\omega t/s$, a particle appears at the detector. Importantly energy of a particle depends linearly on $j$. In other words depending at which moment of time during the period $s2\pi/\omega$ a particle passes close to the detector its energy will be different. Therefore, the potential tilt we have realized is actually a tilt along the time direction. In the quantum description it is best visible in the tight-binding model (cf. Sec.~\ref{stbm}), i.e. the on-site energies change with the temporal index $\alpha$ like $J_{i,\alpha}^{i,\alpha}\approx\alpha2\pi U_x/s$.

\begin{figure}
\includegraphics[width=84mm]{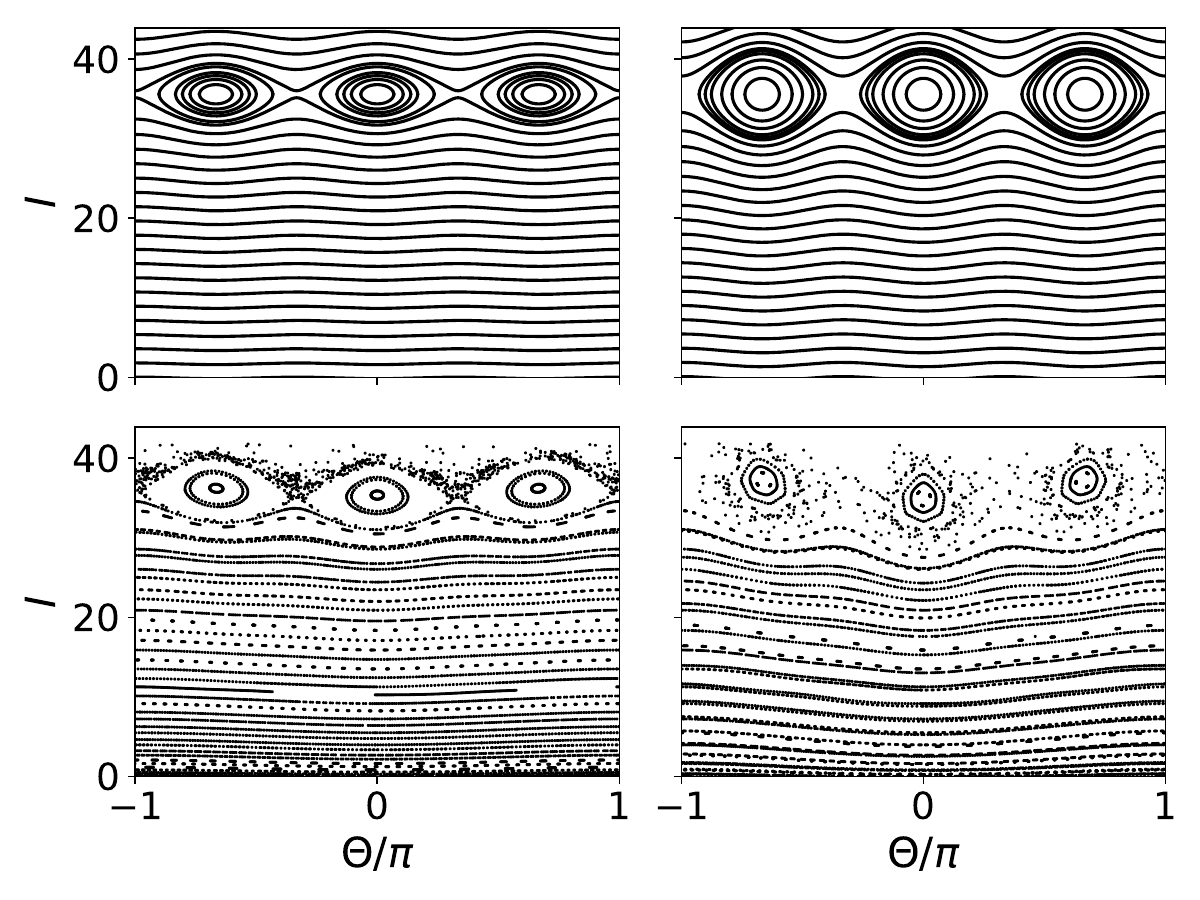}
\caption{Top panels: phase space pictures, in the action-angle variables, generated by the classical secular Hamiltonian (\ref{Sclass_sec}) for $s = 3$, $V_0 = 4320$, $\omega = 240$ and $\lambda = 0.01$ (left) and $\lambda = 0.025$ (right). Bottom panels: stroboscopic maps obtained by numerical integration of the exact equations of motion --- bottom left (right) panel corresponds to the phase space pictures presented in top left (right) panel. The stroboscopic maps are obtained by plotting points $\{\Theta(nT),I(nT)\}$ where $T=2\pi/\omega$ is the period of the lattice shaking and $n$'s are integer.}
\label{Sone}   
\end{figure}

The validity of the secular Hamiltonian (\ref{Sclass_sec}) [or (\ref{Sclass_sec_bar})] can be examined by comparison of the phase space picture it generates with the stroboscopic map obtained from numerical integration of the exact equations of motion. The depth of the optical lattice potential $V_0$ can be arbitrary because it defines the unperturbed part of the Hamiltonian but the strength of the perturbation (determined by $\lambda$) has to be sufficiently small. Figure~\ref{Sone} shows pictures of the classical phase space and indicates that for sufficiently weak shaking the secular approximation method leads to accurate quantitative description of the system. 

\section{Quantum secular approximation method}

Quantization of the classical secular Hamiltonian (\ref{Sclass_sec}) [or (\ref{Sclass_sec_bar})] allows one to obtain quantum description of resonant motion of a particle in a single site of the optical lattice. To incorporate also tunneling transitions between different real-space sites of the optical lattice potential one can switch to the quantum version of the secular approximation method \cite{sBerman1977,sSacha2017rev,sSachaTC2020} which we introduce in the present section.

The Hamiltonian (\ref{SH_basic}) is periodic in time, $H(x,p_x,t+T) = H(x,p_x,t)$ where $T = 2\pi/\omega$. Thus, we may look for time-periodic Floquet states which are eigenstates of the Floquet Hamiltonian ${\cal H} = H-i\partial_t$ \cite{sShirley1965,sBuchleitner2002}. The Hamiltonian (\ref{SH_basic}) is also periodic in space $H(x+\pi,p_x,t)=H(x,p_x,t)$ and consequently the Floquet states have the form of Bloch waves $e^{ikx}u_{k,\alpha}(x,t)$ where $k$ is a quasi-momentum, $u_{k,\alpha}(x+\pi,t)=u_{k,\alpha}(x,t)=u_{k,\alpha}(x,t+T)$. The wavefunctions $u_{k,\alpha}(x,t)$ fulfill the Floquet eigenvalue equation
\be
\left[H^{(0)}(k)+(p_x+k)\lambda\omega\sin\omega t-i\partial_t\right]u_{k,\alpha}=E_{k,\alpha}u_{k,\alpha},
\label{SfullFloquet}
\ee
where 
\be
H^{(0)}(k)=(p_x+k)^2+V_0\sin^2x,
\ee 
and $E_{k,\alpha}$'s are quasi-energies of the system. The index $\alpha$ labels different Floquet states corresponding to the same quasi-momentum $k$. 

\begin{figure}
\includegraphics[width=84mm]{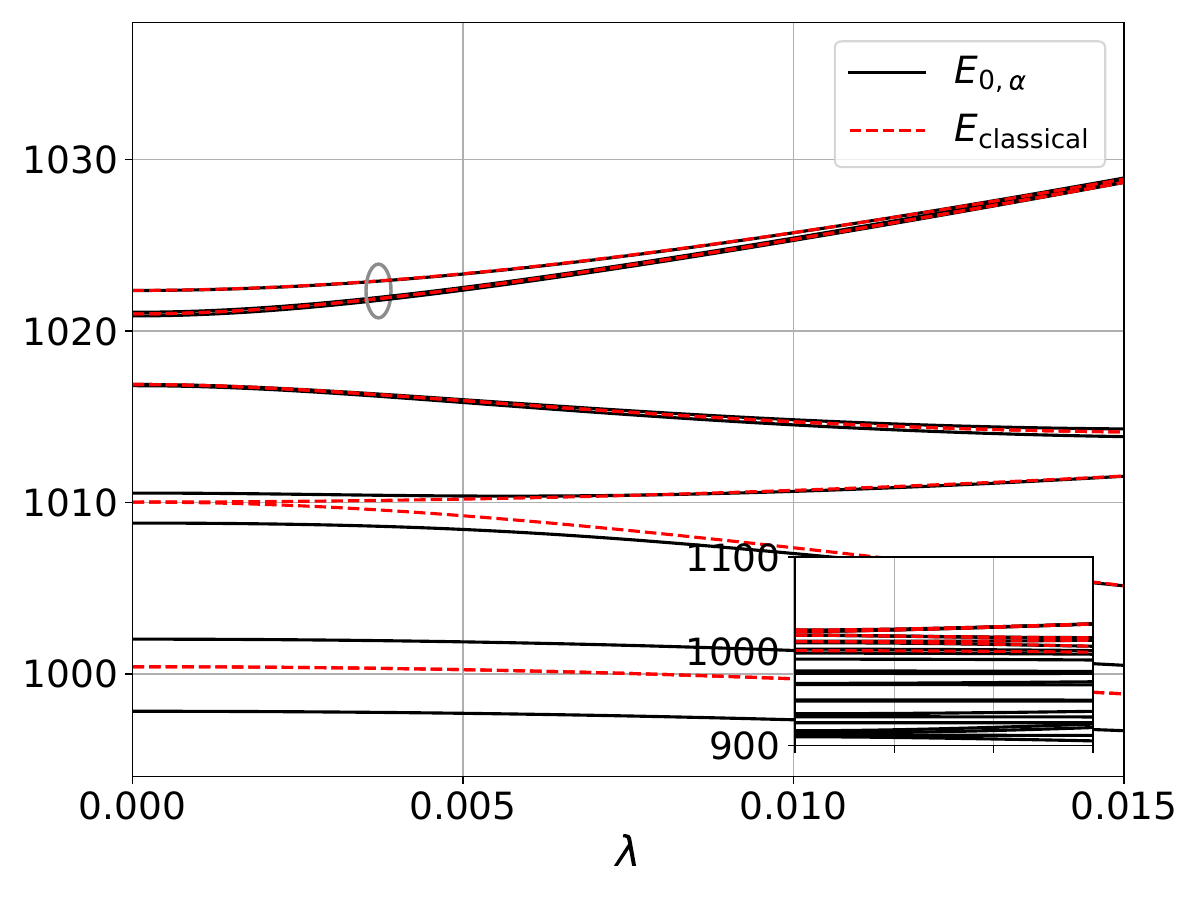}
\caption{Black lines show energy levels obtained by diagonalization of the quantum secular Hamiltonian (\ref{SmatrixH_F}) versus 
$\lambda$ for $s = 3$, $V_0 = 4320$ and $\omega = 240$. Red dashed lines present resonant energy levels obtained by diagonalization of the quantized version of the classical secular Hamiltonian (\ref{Sclass_sec}). Comparison of these two sets of data allows one to identify the resonant energy levels among many eigenenergies of the quantum secular Hamiltonian --- the highest energy levels in the plot (marked by a grey elliptical loop) correspond to the $s=3$ resonant Floquet states we are interested in. Note that only two lines are visible as the lower energy level is degenerate. The inset shows a zoomed out version of the spectrum of the quantum secular Hamiltonian with the vertical energy axis covering a broader range while the range of $\lambda$s is the same.}
\label{Stwo}   
\end{figure}

Solutions of (\ref{SfullFloquet}) related to Floquet states that describe resonant dynamics of a particle can be obtained in a simple way by applying the quantum secular approximation approach. To this end we choose the eigenstates $\psi_{k,n}(x)$ of the unperturbed Hamiltonian,
\be
H^{(0)}(k)\psi_{k,n}(x)=E_{k,n}^{(0)}\psi_{k,n}(x), 
\ee
as the basis for the Hilbert space of a particle. Next
we perform the time-dependent unitary transformation $\psi_{k,n}'=e^{-in\omega t/s}\psi_{k,n}$ which is a quantum analog of the canonical transformation to the frame moving along a resonant orbit. It results in the following matrix elements of the Floquet Hamiltonian 
\bea
\la \psi_{k,n'}'|{\cal H}|\psi_{k,n}'\ra&=&\left(E_{k,n}^{(0)}-n\frac{\omega}{s}+k\lambda\omega \sin\omega t\right)\delta_{n'n}
\cr &&
-\la \psi_{k,n'}|p_x|\psi_{k,n}\ra\frac{i\lambda\omega}{2}
\cr &&
 \times
\left(e^{i(n'-n+s)\frac{\omega}{s}t}-e^{i(n'-n-s)\frac{\omega }{s}t}\right).
\cr &&
\label{SmatrixH_F_full}
\eea
Averaging the above Hamiltonian over time while keeping all quantities fixed (which is valid in the resonant subspace where $|E_{k,n\pm 1}^{(0)}-E_{k,n}^{(0)}|\approx \omega/s$) we get the desired time-independent effective Floquet Hamiltonian 
 \bea
\la \psi_{k,n'}'|{\cal H}|\psi_{k,n}'\ra&\approx&\left(E_{k,n}^{(0)}-n\frac{\omega}{s}\right)\delta_{n'n}
-\la \psi_{k,n'}|p_x|\psi_{k,n}\ra
\cr &&
 \times\frac{i\lambda\omega}{2}
\left(\delta_{n'n-s}-\delta_{n'n+s}\right).
\label{SmatrixH_F}
\eea
Diagonalization of (\ref{SmatrixH_F}) allows us to obtain $s$ resonant Floquet states
\be
e^{ikx}u_{k,\alpha}(x,t)=e^{ikx}\sum_n c_{k,n}^{(\alpha)}e^{-in\omega t/s}\psi_{k,n}(x),
\label{SFloquet_states}
\ee
where $\alpha=1,\dots,s$ and $c_{k,n}^{(\alpha)}$ are constants. To identify the resonant Floquet states among all eigenstates of (\ref{SmatrixH_F}) it is helpful to compare the spectrum of (\ref{SmatrixH_F}) with the spectrum of the quantized version of the classical secular Hamiltonian (\ref{Sclass_sec}). Figure~\ref{Stwo} presents such comparison and demonstrates also consistency and validity of the approximation methods we use. That is, the classical secular Hamiltonian reproduces the exact classical dynamics very well for sufficiently weak time-periodic driving (cf. Fig.~\ref{Sone}) and the spectrum of its quantized version match the resonant energy levels of the quantum secular Hamiltonian (\ref{SmatrixH_F}).

\section{Tight-binding model of time-space crystals}
\label{stbm}

Let us illustrate how to derive a tight-binding model for a time-space crystal by focusing on the system described by the Hamiltonian (\ref{SH_basic}).

We are interested in resonant dynamics of a particle in the shaken optical lattice potential, i.e. we restrict to the Hilbert subspace spanned by the Floquet-Bloch states (\ref{SFloquet_states}). Within this subspace one can define Wannier-like states $w_{i,\alpha}(x,t)$ which are localized wave-packets moving along classical resonant orbits in each $i$-site of the optical lattice potential. That is, in each $i$-site one can define $\alpha=1,\dots,s$ wave-packets which are propagating along the resonant orbit one by one with the period $s2\pi/\omega$ and are delayed with respect to each other by $2\pi/\omega$. 

In practice to obtain $w_{i,\alpha}$ one can diagonalize the position operator $\hat O=x$ (or $\hat O=e^{ix}$) in the subspace spanned by the resonant Floquet-Bloch states, 
\bea
\la k',\alpha'|\hat O|k,\alpha\ra&=&\int dx \left[e^{ik'x}u_{k',\alpha'}(x,t)\right]^*\hat Oe^{ikx}u_{k,\alpha}(x,t).
\cr &&
\label{Sx}
\eea
Any $t$ in (\ref{Sx}) can be chosen, however, moments of time when wave-packets strongly overlap are unfavorable because in this case eigenvalues of $\hat O$ become degenerate and the diagonalization can result in linear combinations of the Wannier states that we look for. Thus we make a specific choice of time $t$ which ensures that the wave-packets are not strongly overlapping. The obtained Wannier states read 
\be
w_{j,\alpha}(x,t)=\sum_{k,\beta}b^{j,\alpha}_{k,\beta}\;e^{ikx}u_{k,\beta}(x,t),
\ee 
where $b^{j,\alpha}_{k,\beta}$ are constants.

Having the Wannier state basis one can derive a tight-binding model by calculating matrix elements of the Floquet Hamiltonian in such a basis, 
\bea
J^{j,\beta}_{i,\alpha}&=&-2\int_0^{sT}\frac{dt}{sT}\la w_{j,\beta}|{\cal H}|w_{i,\alpha}\ra
\cr
&=&-2\sum_{k,\gamma} E_{k,\gamma}\left(b_{k,\gamma}^{j,\beta}\right)^*b_{k,\gamma}^{i,\alpha},
\label{2dJ}
\eea 
where $E_{k,\gamma}$ are quasi-energies, cf. (\ref{SfullFloquet}). Finally we obtain a tight-binding model which corresponds to quasi-energy $\cal E$ of a particle prepared in a state $\psi(x,t)=\sum_{i,\alpha}a_{i,\alpha}w_{i,\alpha}(x,t)$,
\be
{\cal E}=\int_0^{sT}\frac{dt}{sT}\la\psi|{\cal H}|\psi\ra=-\frac12\sum_{i,\alpha,j,\beta}
J^{j,\beta}_{i,\alpha}
a^*_{j,\beta}a_{i,\alpha},
\label{Stb}
\ee
where $J^{j,\beta}_{i,\alpha}$'s have the meaning of tunneling amplitudes and they are dominant for hopping of a particle between nearest neighbor optical lattice sites.

In order to describe $N$ bosonic atoms in the resonantly shaken optical lattice, one may apply the second quantization formalism and obtain the effective Bose-Hubbard Hamiltonian. We are interested in the resonant Hilbert subspace spanned by the single-particle time-dependent basis of the Wannier states $w_{i,\alpha}(x,t)$ and consequently we expand the bosonic field operator in the corresponding annihilation operators
\be
\hat\psi(x,t)\approx\sum_{i,\alpha}w_{i,\alpha}(x,t)\;\hat a_{i,\alpha}, 
\ee
and calculate the second quantized version of the Floquet Hamiltonian
\bea
\hat {\cal H}&=&\int_0^{sT}\frac{dt}{sT}\int dx\;\hat\psi^\dagger(x,t)[H(t)-i\partial_t]\hat\psi(x,t)
\cr &=&-\frac12\sum_{i,\alpha,j,\beta}
J^{j,\beta}_{i,\alpha}
\hat a^\dagger_{j,\beta}\hat a_{i,\alpha}.
\label{HFtbm}
\eea
If contact interactions between ultra-cold atoms are present and they are not strong enough to couple significantly the system to the subspace complementary to the resonant Hilbert space, we may approximate the interaction part of the Floquet Hamiltonian as follows \cite{sSachaTC2020}
\bea
\frac{g_0}{2}\int_0^{sT}\frac{dt}{sT}\int dx\;\hat\psi^\dagger\hat\psi^\dagger\hat\psi\hat\psi&\approx& \frac12\sum_{i',\alpha',j',\beta'}\sum_{i,\alpha,j,\beta}U_{i,\alpha,j,\beta}^{i',\alpha',j',\beta'}
\cr && \times \hat a^\dagger_{j',\beta'}\hat a_{i',\alpha'}^\dagger\hat a_{j,\beta}\hat a_{i,\alpha},
\eea
where 
\be
U_{i,\alpha,j,\beta}^{i',\alpha',j',\beta'}=\int_0^{sT}\frac{dt}{sT}\int dx\;g_0\;w_{i',\alpha'}^*w_{j',\beta'}^*w_{i,\alpha}^*w_{j,\beta}^*.
\label{ssqtbm}
\ee
The on-site interactions $U_{i,\alpha,i,\alpha}^{i,\alpha,i,\alpha}$ are dominant but if the s-wave scattering length of atoms is properly periodically modulated in time, $g_0(t+sT)=g_0(t)$, the long range interactions $U_{i,\alpha,i,\beta}^{i,\alpha,i,\beta}$ between atoms occupying different wavepackets in the same optical lattice sites can be enhanced \cite{sGiergiel2018,sSachaTC2020}. That is, at the moments when two Wannier states $w_{i,\alpha}$ and $w_{i,\beta}$ overlap one may tune the scattering length $g_0(t)$ to a larger value and repeat it with the period $sT$ which results in larger values of $U_{i,\alpha,i,\beta}^{i,\alpha,i,\beta}$ that determine the effective long range interactions, cf. Eq.~(\ref{ssqtbm}).

\section{Six-dimensional crystalline structures}

We have seen that the effective description of a resonantly shaken 1D optical lattice potential leads to the 2D tight-binding model (\ref{Stb}) [or (\ref{HFtbm})]. Now we are going to explain how to obtain a 6D crystalline structure if the 3D separable optical lattice, $V_0(\sin^2x+\sin^2y+\sin^2z)$, is shaken resonantly (with the same frequency $\omega$) along the $x$, $y$ and $z$ directions.

Let us first focus on a single site [denoted by $\vec i=(i_x,i_y,i_z)$] of the 3D optical lattice potential. Due to the fact that the 3D lattice is separable we may define Wannier-like states independently for each degree of freedom, i.e. $w_{i_x,\alpha_x}(x,t)$, $w_{i_y,\alpha_y}(y,t)$ and $w_{i_z,\alpha_z}(z,t)$. It allows us to construct the 3D Wannier states for a particle 
\be
W_{\vec i,\vec \alpha}(\vec r,t) = w_{i_x,\alpha_x}(x,t) \;w_{i_y,\alpha_y}(y,t) \;w_{i_z,\alpha_z}(z,t).
\ee
Thus, in a given site of the 3D optical lattice potential we deal with a space of states labeled by the three-component index $\vec \alpha=(\alpha_x,\alpha_y,\alpha_z)$ where $\alpha_x$, $\alpha_y$, $\alpha_z=1,\dots,s$. In other words we are able to realize a finite 3D crystalline structure in each site of the 3D optical lattice shaken along the three independent directions with the same frequency $\omega$. Such a structure is repeated in each site of the 3D optical lattice and consequently we end up with a 6D crystalline structure whose sites are labeled by two vector indexes $\vec i$ and $\vec \alpha$. The 6D crystalline structure can be described by the tight-binding model (or the Bose-Hubbard Hamiltonian) of the same form as Eq.~(\ref{Stb}) [or (\ref{HFtbm})] but with $i\rightarrow \vec i$ and $\alpha\rightarrow\vec \alpha$. Indeed, if we expand a wavefunction of a particle in the basis of the 3D Wannier states,
\be
\psi(\vec r,t)=\sum_{\vec i,\vec \alpha}W_{\vec i,\vec \alpha}(\vec r,t)\; a_{\vec i,\vec \alpha},
\ee
the quasi-energy of a particle in such a Hilbert subspace reads
\bea
{\cal E}&=&\int_0^{sT}\frac{dt}{sT}\int dx\int dy\int dz\;\psi^*(\vec r,t)
\cr && \times\left[H(x,p_x,t)+H(y,p_y,t)+H(z,p_z,t)-i\partial_t\right]\psi(\vec r,t)
\cr &=&-\frac12\sum_{\vec i,\vec \alpha,\vec j,\vec \beta}
J^{\vec j,\vec \beta}_{\vec i,\vec \alpha}\;
a^*_{\vec j,\vec \beta}\; a_{\vec i,\vec \alpha},
\eea
where
\begin{widetext}
\bea
J^{\vec j,\vec \beta}_{\vec i,\vec \alpha}&=&
-2\int_0^{sT}\frac{dt}{sT}
\left(\int dx \;w^*_{j_x,\beta_x}[H(x,p_x,t)-i\partial_t]w_{i_x,\alpha_x}\right)
\left(\int dy \;w^*_{j_y,\beta_y}w_{i_y,\alpha_y}\right)
\left(\int dz \;w^*_{j_z,\beta_z}w_{i_z,\alpha_z}\right)
\cr && 
-2\int_0^{sT}\frac{dt}{sT}\left(\int dx \;w^*_{j_x,\beta_x}w_{i_x,\alpha_x}\right)
\left(\int dy \;w^*_{j_y,\beta_y}[H(y,p_y,t)-i\partial_t]w_{i_y,\alpha_y}\right)
\left(\int dz \;w^*_{j_z,\beta_z}w_{i_z,\alpha_z}\right)
\cr && 
-2\int_0^{sT}\frac{dt}{sT}\left(\int dx\;w^*_{j_x,\beta_x}w_{i_x,\alpha_x}\right)
\left(\int dy \;w^*_{j_y,\beta_y}w_{i_y,\alpha_y}\right)
\left(\int dz \;w^*_{j_z,\beta_z}[H(z,p_z,t)-i\partial_t]w_{i_z,\alpha_z}\right)
\cr &&
\cr &=& 
J_{i_x,\alpha_x}^{j_x,\beta_x}\;\delta_{\beta_y,\alpha_y}\;\delta_{j_y,i_y}\;\delta_{\beta_z,\alpha_z}\;\delta_{j_z,i_z}
+\delta_{\beta_x,\alpha_x}\;\delta_{j_x,i_x}\;J_{i_y,\alpha_y}^{j_y,\beta_y}\;\delta_{\beta_z,\alpha_z}\;\delta_{j_z,i_z}
+\delta_{\beta_x,\alpha_x}\;\delta_{j_x,i_x}\;\delta_{\beta_y,\alpha_y}\;\delta_{j_y,i_y}\;J_{i_z,\alpha_z}^{j_z,\beta_z},
\cr &&
\label{6dJ}
\eea
\end{widetext}
where $H$ is given in Eq.~(\ref{SH_basic}) [or Eq.~(\ref{SH_basic_bar})] and $J_{i,\alpha}^{j,\beta}$ are tunneling rates obtained in the 2D case, see Eq.~(\ref{2dJ}). The Kronecker deltas in Eq.~(\ref{6dJ}) indicate the fact that the 6D lattice is separable into three independent 2D lattices. Note that the Floquet-Bloch states and Wannier states form orthogonal bases at any time \cite{sShirley1965,sBuchleitner2002} therefore the integrals like 
\bea
 \int dx \;w^*_{j_x,\beta_x}(x,t)\;w_{i_x,\alpha_x}(x,t)&=&
\delta_{\beta_x,\alpha_x}\;\delta_{j_x,i_x}
\eea
are time-independent.




\end{document}